\documentclass[12pt]{article}

\usepackage{amsmath}%
\usepackage{amsfonts}%
\usepackage{amssymb}%
\usepackage{graphicx}

\input amssym.def
\input amssym
\newcommand{\be}{\begin{equation}}
\newcommand{\ee}{\end{equation}}
\begin{document}

\renewcommand{\theequation}{\arabic{section}.\arabic{equation}}

\title{Topological gravity on plumbed V-cobordisms}
\author{Vladimir N. Efremov\thanks{Mathematics Department, CUCEI,
University of Guadalajara, Guadalajara, Mexico.}
\thanks{E-mail:
efremov@udgserv.cencar.udg.mx}\\
Nikolai V. Mitskievich\thanks{Physics Department, CUCEI,
University of Guadalajara, Guadalajara, Mexico.}
\thanks{E-mail:
mitskievich03@yahoo.com.mx}\\
Alfonso M. Hern\'andez Magdaleno$^\ddag$\\
and\\
Ramona Serrano Bautista$^\ast$
}
\date{~}

\maketitle

%\newpage

\begin{abstract}
An ensemble of cosmological models based on generalized
$BF$-theory is constructed where the r{\^o}le of vacuum (zero-level)
coupling constants is played by topologically invariant rational
intersection forms (cosmological-constant matrices) of 4-dimensional
plumbed V-cobordisms which are interpreted as Euclidean spacetime 
regions. For these regions describing topology changes, the
rational and integer intersection matrices are calculated. A
relation is found between the hierarchy of certain elements of
these matrices and the hierarchy of coupling constants of
the universal (low-energy) interactions.\\

\end{abstract}

\newpage
\section{Introduction}  \label{s1}

The principal motivation for the present paper is to
relate some results in low dimensional topology to $BF$-theory
\cite{[BlauThom],[Hor]} also known as topological gravity \cite{[BaezMG]}.
Properly speaking, we are trying to realize an analogue
of the classical Abelian $BF$-model on plumbed (graph)
V-manifolds and V-cobordisms 
\cite{[Stake],[Sieb],[Orlik],[EisNeum],[FuFuUe],[SavFF]}. 
In addition to the usual characteristics of
topological field theories (absence of local degrees of
freedom, finite dimensionality of phase space, {\it etc}.),
in our model as analogues of cosmological (coupling)
constants there appear rational intersection forms
(matrices) which are the basic topological invariants
of plumbed V-manifolds \cite{[FintSt],[FuFuUe],[SavFF]}. Thus in
our model is realized the idea that the primary values of
coupling constants (which correspond to a vacuum without
any local excitations) are to be looked for at the
topological level (see {\it e.g.} \cite{[KlSuBa]}).

The paper is organized as follows. Section 2 starts
with a short review of the main ideas concerning plumbed
V-cobordisms and V-manifolds. We reproduce definitions of
basic topological invariants of these spaces, specifically,
the rational and integer intersection matrices. We also
formulate the important result \cite{[FuFuUe]} concerning 
reciprocality (mutual inverse property) of these matrices
written in natural bases.

Section 3 is dedicated to construction of Abelian
$BFE$-models on plumbed V-cobordisms. We show that these
models possess an essentially new property in comparison
with usual $BF$-theories: the r\^{o}le of vacuum (zero-level)
coupling constants is played by the topologically invariant
intersection forms of plumbed V-cobordisms.

In section 4 we give a collection of non-trivial examples
which realize the ideas proposed in section 3. As a starting
point is taken the primary sequence of 3-dimensional Seifert
fibered homology spheres (Sfh-spheres) characterized by Seifert
invariants constructed of first nine prime numbers $p_1=2$,
$p_2=3, \dots, p_9=23$. With the help of introduced by us
derivative operation acting on Sfh-spheres, we construct
the basic two-parametric family of Sfh-spheres 
$\Sigma^{(l)}_{\phantom{|}n}$ which by means of the well-known
splicing operation are pasted together into ${\Bbb Z}$-homology
spheres $M^{(l)}_{\phantom{|}n}$. These latter ones turn out
to be the boundary components of 4-dimensional plumbed
V-cobordisms ${X_D}^{(l)}_{\phantom{|}n}$ together with
a specific collection of lens spaces. Just on this set of
cobordisms we are constructing the respective $BFE$-systems
$S^{(l)}_{\phantom{|}n}$ with cosmological coupling matrices
being, as we have proven, the representatives of rational
intersection forms of ${X_D}^{(l)}_{\phantom{|}n}$. Namely
these V-cobordisms can be considered as construction units
in gluing together the compound V-cobordisms which describe
topology changes $M^{(l)}_{\phantom{|}n}\longrightarrow
M^{(l)}_{\phantom{|}n+1}$ between ${\Bbb Z}$-homology spheres
(this is sketched in Conclusions, section 5).

Further we give results of calculation of both rational and
integer intersection matrices for a finite ensemble of
plumbed V-cobordisms ${X_D}^{(l)}_{\phantom{|}n}$ with
$n,l=0,\dots,4$. So we find a relation between the
diagonal elements of the integer intersection matrices and
Euler numbers of the basic family of Sfh-spheres
$\Sigma^{(l)}_{\phantom{|}n}$. These Euler numbers fairly
well reproduce the hierarchy of the dimensionless low-energy
coupling constants.

The standard notations ${\Bbb Z}$, ${\Bbb R}$ and ${\Bbb C}$ are
used for the sets of integer, real and complex numbers, respectively.

\section{Plumbed 4- and 3-dimensional manifolds}\label{s2}

In order to build a new version of the Abelian $BF$ theory
on 4-dimensional pV-manifolds and pV-cobordisms whose boundaries
are ${\Bbb Z}$-homology spheres and lens spaces, we first give
some necessary definitions essentially following the works of
Saveliev \cite{[SavFF],[Sav2002]}. These ideas can be traced back to
\cite{[Sieb],[Orlik],[NeumRaym],[EisNeum]}.

\subsection{Plumbing and splicing operations}\label{s2.1}

A {\it plumbing graph} $\Gamma$ is a graph with no cycles (a finite
tree) each of whose vertices $v_i$ carries an integer weight
$e_i, ~ i=1,\dots,s$. The $D^2$-bundle $Y(e_i)$ is associated to
each vertex $v_i$, whose Euler class (self-intersection number of
zero-section) is $e_i$. If the vertex $v_i$ has $d_i$ edges
connected to it on the graph $\Gamma$, choose $d_i$ disjoint
discs in the base $S^2$ of $Y(e_i)$ and call the disc bundle over
the $j$th disc $B_{ij}=\left(D^2_j\times D^2\right)_i$. When two
vertices $v_i$ and $v_k$ are connected by an edge, the disc
bundles $B_{ij}$ and $B_{kl}$ should be identified by exchanging
the base and fiber coordinates \cite{[Orlik]}. This pasting operation is
called {\it plumbing}, and the resulting smooth 4-manifold
$P(\Gamma)$ is known as {\it plumbed 4-manifold}. Its boundary
$M(\Gamma)=\partial P(\Gamma)$ is referred to as a {\it plumbed
3-manifold}.

Since the homology group $H_1(P(\Gamma),{\Bbb Z})=0$, the unique
non-trivial homology characteristic is $H_2(P(\Gamma),{\Bbb Z})$
which has a natural basis (set of generators) represented by the
zero-sections of the plumbed bundles. All these sections are
embedded 2-spheres $z_a\mbox{ where }a=1,\dots,r=
\mbox{ rank }H_2(P(\Gamma),{\Bbb Z})$, and they
can be oriented in such a way that the intersection (bilinear)
form \cite{[Sav2002]}
\be A: ~H_2(P(\Gamma),{\Bbb Z})\otimes H_2(P(\Gamma),{\Bbb Z})
\rightarrow {\Bbb Z} \label{2.1} \ee
will be represented by the $r\times r$-matrix $A(\Gamma)=
(a_{ij})$ with the entries: $a_{ij}=e_i$ if $i=j$; $a_{ij}=1$ if
the vertex $v_i$ is connected to $v_j$ by an edge; and $a_{ij}=0$
otherwise.

Let $M$ be a Seifert fibered 3-manifold over $S^2$ with
unnormalized Seifert invariants \cite{[NeumRaym]} $(a_i,b_i), ~i=1,
\dots,n, ~a_i>1$. It can be obtained as the boundary of the
plumbed 4-manifold $P(\Gamma)$ where $\Gamma$ is a star
graph shown in figure 1 \cite{[Orlik]}.

\begin{figure}[h]     %[tbh]
%\vspace*{-4.3cm}
\begin{center}
\setlength{\unitlength}{1pt}
\begin{picture}(150,80)
\put(29,60){\line(1,0){65}} \put(29,00){\line(1,0){65}}
\put(115,60){\line(1,0){25}} \put(115,00){\line(1,0){25}}
\put(0,30){\line(1,1){30}} \put(0,30){\line(1,-1){30}}
\put(30,59.5){\makebox(.5,.5){$\bullet$}}
\put(30,-.5){\makebox(.5,.5){$\bullet$}}
\put(95,57){\makebox(20,5){...}} \put(95,-3){\makebox(20,5){...}}
\put(140,59.5){\makebox(.5,.5){$\bullet$}}
\put(140,-.5){\makebox(.5,.5){$\bullet$}}
\put(70,59.5){\makebox(.5,.5){$\bullet$}}
\put(70,-.5){\makebox(.5,.5){$\bullet$}}
\put(0,29.5){\makebox(.5,.5){$\bullet$}}
\put(-9,29.5){\makebox(.5,.5){$0$}}
\put(30,69.5){\makebox(.5,.5){$t_{11}$}}
\put(30,-10.5){\makebox(.5,.5){$t_{n1}$}}
\put(140,69.5){\makebox(.5,.5){$t_{1m_1}$}}
\put(140,-10.5){\makebox(.5,.5){$t_{nm_n}$}}
\put(70,69.5){\makebox(.5,.5){$t_{12}$}}
\put(70,-10.5){\makebox(.5,.5){$t_{n2}$}}
\put(55,12){\makebox(20,5){...}} \put(95,12){\makebox(20,5){...}}
\put(55,27){\makebox(20,5){...}} \put(95,27){\makebox(20,5){...}}
\put(55,42){\makebox(20,5){...}} \put(95,42){\makebox(20,5){...}}
\end{picture}\\
\vspace*{1.cm} {\footnotesize Figure 1.}
\end{center}
\end{figure}

The integer weights $t_{ij}$ in this graph are found from
continued fractions $a_i/b_i=[t_{i1},\dots,t_{im_i}]$; here
$$
[t_1,\dots,t_k]=t_1-\frac{1}{t_2-\displaystyle\frac{1}{\cdots-
\displaystyle\frac{1}{t_k}}}.
$$
Note that among closed 3-manifolds ${\Bbb Z}$-homology spheres
$M$ are characterized by $H_1(M,{\Bbb Z})=0$.

A Seifert fibered 3-manifold $M$ is a ${\Bbb Z}$-homology
sphere (Sfh-sphere) iff the determinant of matrix $A(\Gamma)$
associated with the plumbing graph in figure 1, is $\pm 1$ which
is equivalent to the condition
\be\label{2.2} a\sum b_i/a_i=\pm1 \ee
where $a=a_1\cdots a_n$. We fix an orientation on $M$ by
choosing $+1$ in (\ref{2.2}). From (\ref{2.2}) it follows that
$b_i\sigma_i=1\mbox{ mod }a_i$ where $\sigma_i=a/a_i$; thus
the set of $a_i$ proves to be sufficient for the complete
determination of any Sfh-sphere. This is why we use the standard
notation $\Sigma(a_1,\dots,a_n)$ for Sfh-spheres.

Lens spaces represent another case of Seifert fibered manifolds.
Expanding $-p/q=[t_1,\dots,t_n]$ into a continued fraction, we
encounter $L(p,q)$ as a boundary of the 4-manifold obtained by
plumbing on the chain $\Gamma$ shown in Figure 2.

\begin{figure}[h]     %[tbh]
%\vspace*{-4.3cm}
\begin{center}
\setlength{\unitlength}{1pt}
\begin{picture}(150,15)
\put(2,00){\line(1,0){89}} \put(117,00){\line(1,0){23}}
\put(3,-.5){\makebox(.5,.5){$\bullet$}}
\put(95,-3){\makebox(20,5){...}}
\put(140,-.5){\makebox(.5,.5){$\bullet$}}
\put(50,-.5){\makebox(.5,.5){$\bullet$}}
\put(4.5,-10.5){\makebox(.5,.5){$t_{1}$}}
\put(141.5,-10.5){\makebox(.5,.5){$t_{n}$}}
\put(51.5,-10.5){\makebox(.5,.5){$t_{2}$}}
\end{picture}\\
\vspace*{1.cm} {\footnotesize Figure 2.}
\end{center}
\end{figure}

Notice that this plumbing graph simultaneously represents the
lens space $L(p,q^*)$ with $-p/q^*=[t_n,\dots,t_1]$ where
$qq^*=1$ mod $p$. This reflects the fact that $L(p,q)$ and
$L(p,q^*)$ are homeomorphic.

In section 4 we shall build universe models which are plumbed 
4-manifolds (cobordisms) with boundaries whose components are
homeomorphic to lens spaces or to ${\Bbb Z}$-homology spheres,
the latter ones being constructed by splicing of certain set
of Sfh-spheres with three exceptional fibers, {\it i.e.} 
$\Sigma(a_1,a_2,a_3)$. Therefore it is worth giving now the
general definition of the splicing operation as well as the
plumbed V-cobordism.

By a link \cite{[EisNeum]} $(\Sigma,K)=(\Sigma,S_1\bigcup\cdots 
\bigcup S_m)$
we mean a pair consisting of oriented ${\Bbb Z}$-homology sphere
$\Sigma$ and a collection $K$ of disjoint oriented knots
$S_1,\dots,S_m$ in $\Sigma$. Empty links are just ${\Bbb 
Z}$-homology spheres. Note that the links $(S^3,K)$ where $S^3$
is an ordinary 3-sphere, are also allowed. If the link
components $S_1,\dots,S_m$ are fibers in $\Sigma$, then the link
$(\Sigma,K)$ is called a Seifert link. Let $(\Sigma,K)$ and
$(\Sigma',K')$ be links and choose components $S\in K$ and
$S'\in K'$. Let also $N(S)$ and $N(S')$ be their tubular
neighbourhoods, while $m,l\subset\partial N(S)$ and
$m',l'\subset\partial N(S')$ be standard meridians and longitudes.
The manifold $\Sigma''=(\Sigma\setminus\textnormal{int}N(S))
\bigcup(\Sigma'\setminus\textnormal{int}N(S'))$ obtained by
pasting along the torus boundaries by matching $m$ to $l'$ and
$m'$ to $l$, is a ${\Bbb Z}$-homology sphere. The link $\left(
\Sigma'',(K\setminus S)\bigcup(K'\setminus S')\right)$ is called
the splice (splicing) of $(\Sigma,K)$ and $(\Sigma',K')$ along
$S$ and $S'$. We shall use the standard notation $\Sigma''=
\Sigma\frac{~}{S ~ S'}\Sigma'$ or simply $\Sigma''=
\Sigma\frac{~ ~}{~ ~ ~}\Sigma'$. Any link which can be obtained
from a finite number of Seifert links by splicing is called a
graph link. Empty graph links are precisely the plumbed (graph)
${\Bbb Z}$-homology spheres.

A plumbed graph $\Gamma$ with added arrowhead vertices,
denoted as $\overline{\Gamma}$, represents a link $K$ in a
(plumbed) 3-manifold $M=M(\overline{\Gamma})$ as follows: Let
$\Gamma$ be $\overline{\Gamma}$ with all the arrows deleted,
and put $M=\partial P(\Gamma)$. Each arrowhead vertex $v_j$
of $\overline{\Gamma}$ is attached at some vertex of $\Gamma$,
and to this arrow we associate a fiber $S_j$ of the bundle
$Y(e_i)$ used in the plumbing \cite{[EisNeum],[SavFF]}.

The splicing operation on graph links can be described in
terms of plumbing graphs as follows. Suppose that two graph 
links are represented by their plumbing diagrams $\overline{\Gamma}$
and $\overline{\Gamma'}$ (see Figure 3) with arrows attached to vertices
$e_n$ and $e'_m$, respectively.
The corresponding plumbing diagram for a spliced link is
shown in Figure 4
where $a=\det A(\Gamma_0)/\det A(\Gamma)$, while $\Gamma$ is
the plumbing graph $\overline{\Gamma}$ with the arrow deleted,
and $\Gamma_0$ is a portion of $\Gamma$ obtained by removing
the $n$th vertex weighted by $e_n$ as well as all its adjacent
edges. Another integer $a'$ is similarly obtained from the
graph $\Gamma'$ (examples see in \cite{[EisNeum],[SavFF]}). The above
description of splicing in terms of plumbing graphs makes it
possible to treat splicing as an operation on the corresponding
plumbed 4-manifold; moreover, $M(\overline{\Gamma})=\partial 
P(\Gamma)$.

\begin{figure}[h]     %[tbh]
%\vspace*{-4.3cm}
\begin{center}
\setlength{\unitlength}{1pt}
\begin{picture}(150,50)
\put(2,00){\line(1,0){45}} \put(140,00){\line(-1,0){45}}
\multiput(2,-0.5)(-3,-3){10}{.}   %{\line(-1,-1){10}}
\multiput(2,-0.5)(-3,3){10}{.}     %{\line(-1,1){10}}
\multiput(140,0)(3,3){10}{.} \multiput(140,0)(3,-3){10}{.}
\put(47,0){\vector(1,0){5}} \put(95,0){\vector(-1,0){5}}
\put(5,-15){\makebox(10,5){$e_n$}}
\put(135,-15){\makebox(10,5){$e'_m$}}
\put(140,-.5){\makebox(.5,.5){$\bullet$}}
\put(2,-.5){\makebox(.5,.5){$\bullet$}}
\end{picture}\\
\vspace*{1.cm} {\footnotesize Figure 3.}
\end{center}
\end{figure}

\begin{figure}[h]     %[tbh]
%\vspace*{-4.3cm}
\begin{center}
\setlength{\unitlength}{1pt}
\begin{picture}(150,50)
\put(2,00){\line(1,0){140}} %\put(140,00){\line(-1,0){45}}
\multiput(2,-0.5)(-3,-3){10}{.}   %{\line(-1,-1){10}}
\multiput(2,-0.5)(-3,3){10}{.}     %{\line(-1,1){10}}
\multiput(140,0)(3,3){10}{.} \multiput(140,0)(3,-3){10}{.}
\put(47,-0.5){\makebox(.5,.5){$\bullet$}}
\put(95,-0.5){\makebox(.5,.5){$\bullet$}}
\put(5,-15){\makebox(10,5){$e_n$}}
\put(135,-15){\makebox(10,5){$e'_m$}}
\put(140,-.5){\makebox(.5,.5){$\bullet$}}
\put(2,-.5){\makebox(.5,.5){$\bullet$}}
\put(47,-15){\makebox(.5,.5){$a$}}
\put(97,-15){\makebox(.5,.5){$a'$}}
\end{picture}\\
\vspace*{1.cm} {\footnotesize Figure 4.}
\end{center}
\end{figure}

\subsection{Plumbed V-cobordisms and V-manifolds}\label{s2.2}

A {\it plumbed V-cobordism} (pV-cobordism) is called plumbed
4-manifold when it is a cobordism between a plumbed 3-manifold
and a disjoint union of lens spaces. pV-cobordisms may be
considered as models of elementary topology changes. They are
constructed as follows.

Let $P(\Gamma)$ be a plumbed 4-manifold corresponding to graph
$\Gamma$, and $\Gamma^{\rm ch}$ be a chain in $\Gamma$ of the
form shown in Figure 2. Plumbing on $\Gamma^{\rm ch}$ yields a
submanifold $P(\Gamma^{\rm ch})$ of $P(\Gamma)$ whose boundary is
a lens space $L(p,q)$. The closure of $P(\Gamma)\setminus
P(\Gamma^{\rm ch})$ is a smooth compact 4-manifold $X_D$ with
oriented boundary $-L(p,q)\bigsqcup\partial P(\Gamma)$ \cite{[SavFF]}
where $\bigsqcup$ denotes the disjoint sum operation.
Starting with several chains $\Gamma^{\rm ch}_i$ ($i=\overline{1,I}$)
in $\Gamma$ (where $\overline{0,I}$ is the integer
numbers interval from 0 to $I$), one can introduce a cobordism 
between $\partial
P(\Gamma)$ and the disjoint union $L=\displaystyle
\bigsqcup_{i=1}^I L(p_i,q_i)$ of several lens spaces. The chains
$\left\{\Gamma^{\rm ch}_i\right\}$ must be disjoint in the following
sense: no two chains should have a common vertex, and no edges of
$\Gamma$ should have one endpoint on one chain and another, on 
another one. Such cobordisms will be called pV-cobordisms. A
pV-cobordism is always a smooth manifold. The notation `V-' refers
to the fact that each lens space $L(p_i,q_i)$ on the boundary of
$X_D$ may be eliminated by pasting a cone $cL(p_i,q_i)$ over the
lens space. This yields the well known V-manifold \cite{[Stake]}
\be \label{2.3} 
X=X_D\bigsqcup_{i=1}^I cL(p_i,q_i) 
\ee 
with isolated singular points. V-manifolds of the type $X$ are
called {\it plumbed V-manifolds} (pV-manifolds).

A pV-cobordism $X_D$ can be adequately represented by the so-called
{\it decorated plumbing graph} $\Gamma_D$ \cite{[SavFF]} (this explains
the subindex $_D$). Such graphs are decorated with extra ovals
(or circles), each enclosing exactly one chain $\Gamma^{\rm ch}_i$
(or vertex as a particular case of chain). (The above conditions on
the chains $\Gamma^{\rm ch}_i$ to be disjoint translates into the
following conditions on the decorating ovals: Any two 2-disks
bounded by decorating ovals are disjoint, and no edge of $\Gamma_D$
is intersected by more than one decorating oval.) Examples are
considered in section 4.

\subsection{Integral and rational intersection forms of
pV-co\-bordisms}\label{s2.3}

Let $X_D$ be a pV-cobordism with the boundary
$$
\partial X_D=-\bigsqcup^I_{i=1}L(p_i,q_i)\bigsqcup M
$$
where $M$ is a ${\Bbb Z}$-homology sphere. In this case $H^1(X_D,
{\Bbb Z})=0$, and there exists the exact sequence \cite{[FintSt],
[Sav99],[Span]}:
\be \label{2.4}
0\rightarrow H^2(X_D,\partial X_D,{\Bbb Z})\stackrel{j^*}{\rightarrow}
H^2(X_D,{\Bbb Z})\stackrel{i^*}{\rightarrow}
H^2(\partial X_D,{\Bbb Z})=\bigoplus^I_{i=1}{\Bbb Z}_{p_i}\rightarrow 0.
\ee
Note that the cohomology group $H^2(X_D,\partial X_D,{\Bbb Z})$
is isomorphic to the $H^2(X,{\Bbb Z})$ where the pV-manifold is
defined in (\ref{2.3}).

As a consequence of the Poincar\'e--Lefschetz duality, the
integral intersection form can be defined
\be \label{2.5}
\omega_{{\Bbb Z}}:H^2(X_D,\partial X_D,{\Bbb Z})\otimes
H^2(X_D,{\Bbb Z})\rightarrow{\Bbb Z}
\ee
by means of the cup product pairing \cite{[Sav99]}
\be \label{2.6}
\omega_{{\Bbb Z}}(b,e)=\left<b\cup e,[X_D,\partial X_D]\right>
\ee
for each $b\in H^2(X_D,\partial X_D,{\Bbb Z})$ and $e\in
H^2(X_D,{\Bbb Z})$. The operation $\cup$ is known as cup product,
and $[X_D,\partial X_D]$ is relative fundamental class 
\cite{[Span],[FintSt]}. 
Note that the intersection form $\omega_{{\Bbb Z}}$
becomes non-degenerate after factoring out the torsion
subgroup Tor$\left(H^2(X_D,{\Bbb Z})\right)$ \cite{[Sav99]}.

From the exactness of the sequence (\ref{2.4}) and since
$H^2(\partial X_D,{\Bbb Z})$ is a pure torsion, for any $e'\in
H^2(X_D,{\Bbb Z})$ there exists $k\in{\Bbb Z}$ such that
$i^*(ke')=0$. Hence $ke'=j^*(b)$ for the unique $b\in
H^2(X_D,\partial X_D,{\Bbb Z})$. Therefore it is natural to
define the rational intersection form \cite{[FintSt],[FuFuUe]}
\be \label{2.7}
\omega_{{\Bbb Q}}(e',e)=\left<e'\cup e,[X_D,\partial X_D]\right>
:=\frac{1}{k}\left<b\cup e,[X_D,\partial X_D]\right>
\ee
for any pair $e,e'\in H^2(X_D,{\Bbb Z})$.

We shall use the cohomological version of the Proposition 4
from \cite{[FuFuUe]} which can be formulated as follows: Let $X_D$
be a pV-cobordism with $H^1(X_D,{\Bbb Z})=0$. If we choose
a basis $b_i$ of $H^2(X_D,\partial X_D,{\Bbb Z})$ and a dual
basis $e^i$ of $H^2(X_D,{\Bbb Z})$ ($i=1,\dots,r, ~ r={\rm rank }
H^2(X_D,{\Bbb Z})$), dual in the sense of 
\be \label{2.8}
\omega_{{\Bbb Z}}
(b_i,e^j)=\delta_i^j,
\ee
then the integral intersection matrix
$g_{ij}=\omega_{{\Bbb Z}}(b_i,b_j)$ for $H^2(X_D,\partial X_D,
{\Bbb Z})$ is inverse of the rational intersection matrix
$\lambda^{ij}=\omega_{{\Bbb Q}}(e^i,e^j)$ for $H^2(X_D,{\Bbb Z})$.

It is important that the just mentioned intersection forms
(matrices) are the basic topological invariants of compact
4-manifolds. In the only cases being here under consideration, 
namely those of
pV-cobordisms (pV-manifolds) with vanishing first cohomology
groups, all (co)homology information about these pV-cobordisms
(pV-manifolds) is contained in the second (co)homology groups,
and hence also in the intersection matrices $g_{ij}$ and
$\lambda^{ij}$. These matrices are easily calculated by means
of the procedure described in detail in \cite{[SavFF]}, see also
\cite{[EisNeum]}.

\section{Abelian $BFE$-theory on pV-cobordisms}\label{s3}
\setcounter{equation}{0}

In this section we propose a generalized version of topological
$BF$-type theory (known as 4-dimensional topological gravity)
on pV-cobordisms and pV-manifolds.\footnote{Due to the group
isomorphism $H^2(X_D,\partial X_D,{\Bbb Z})\cong
H^2(X,{\Bbb Z})$ (see subsection \ref{s2.3}), the model we
propose below can be constructed in the same way on
pV-cobordisms and pV-manifolds. Thus we restrict ourselves to
pV-cobordisms
only.} The generalization is related to the fact that ranks of
the 2-cohomology groups of pV-cobordisms are in general greater
than one, {\it i.e.} rank$\,H^2(X_D,\partial X_D,{\Bbb Z})=$ rank$
\,H^2(X_D,{\Bbb Z})=r\geq1$. Therefore the set of basic fields
should consist of $r$   (Abelian) forms $B_a, 
~a=1,\dots,r$, as well as $r$ 2-forms $E^a$ dual to $B_a$.

This model which we call $BFE$-theory, will have the following
important properties:

(1) The r{\^o}le of vacuum (zero-level) coupling constants is
played by the topologically invariant intersection forms
(cosmological-constant matrices).

(2) The space ${\cal N}$ of classical solutions (phase space)
of $BFE$-system will be a finite-dimensional vector space
\be \label{3.1}
{\cal N}=H^2(X_D,{\Bbb R})\oplus H^2(X_D,{\Bbb R})
\ee
where $H^2(X_D,{\Bbb R})$ is de Rham-2-cohomology group of
pV-cobordism $X_D$. (The possibility to define the discrete
phase space
\be \label{3.2}
{\cal N}_{\rm discr}=H^2(X_D,{\Bbb Z})\oplus 
H^2(X_D,\partial X_D,{\Bbb Z})
\ee
will be considered elsewhere.)

To realize a model with properties (1) and (2), we introduce
a family of linearly independent elements $B_a$ of the cochain
complex $C^2(X_D,\partial X_D,{\Bbb Z})\otimes{\Bbb R}$. These
elements (cochains) are represented by a set of Abelian 2-forms
which we also denote as $B_a$. The equations of motion (\ref{3.9})
and the gauge symmetries (\ref{3.8}) of our model yield (see
below) the property of 2-forms $B_a$ to be closed and defined up to
exact forms $d\chi_a$. Thus we can assume that $B_a$ form a basis
of the group $H^2(X_D,\partial X_D,{\Bbb Z})\otimes{\Bbb R}$.
Therefore the index $a$ runs from 1 to $r=$ rank
\,$\,H^2(X_D,\partial X_D,{\Bbb Z})\otimes{\Bbb R}$. Note that the
number $r$ of the basis elements $B_a$ is also equal to the number
of vertices exterior to decorated chains $\Gamma^{\rm ch}_i$ of
the decorated graph $\Gamma_D$ which determines the pV-cobordism
$X_D$ \cite{[SavFF]}, see also examples in section 4 and the
Observation 3 below.

Let us introduce a suitable set of Abelian 1-forms $A^a, ~a=1,\dots,
r$, and a constant non-degenerate symmetric $r\times r$-matrix
$\Lambda^{ab}$. Then it is natural to write the action
\be \label{3.3}
S^{(r)}_{BF}=\int_{X_D}\left\{B_a\wedge F^a+\frac{1}{2}
\Lambda^{ab}B_a\wedge B_b\right\}
\ee
which is a direct generalization of the ordinary Abelian 
$BF$-theory action (for $r=1$ and with the $\Lambda$ term or
cosmological constant) \cite{[BlauThom],[Hor],[Baez]}
\be \label{3.4}
S^{(1)}_{BF}=\int_{X_D}\left\{B\wedge F+\frac{\Lambda}{2}
B\wedge B\right\}.
\ee
In (\ref{3.3}), $F^a={\mathrm d}A^a$, d is the exterior derivative,
and in repeated indices the summation convention is applied.
The constant matrix $\Lambda^{ab}$ occupies the place of
cosmological constant, thus we can call it either 
{\it `cosmological constant matrix'} or 
{\it `coupling constant matrix'}.\\

\noindent\underline{Observation 1.} The action (\ref{3.3}) has
the appearance the non-Abelian $BF$-theory action
$$
S_{\rm GR}=\int\left\{B^i\wedge F^i
+\phi_{ij}B^i\wedge B^j+\frac{\Lambda}{2}
B^i\wedge B^i\right\}
$$
which is equivalent to general relativity 
\cite{[Pleb],[CapDJ],[ReisRav]}.                    
While in $S_{\rm GR}$ the traceless symmetric 0-form $\phi_{ij}$
is a Lagrangian multiplier, our constant matrix $\Lambda^{ab}$,
as it is shown below, coincides with the rational intersection
form and is the basic topological invariant of the pV-cobordism
$X_D$. A comparison of (\ref{3.3}) and $S_{\rm GR}$ shows that
an immediate analogue of the cosmological constant $\Lambda$ is
trace of $\Lambda^{ab}$. Moreover, the indices used in
these two theories have different nature: in our model, $a,b$
enumerate Abelian fields, and in the non-Abelian $BF$-theory,
$i,j$ are the gauge group indices.\\

However the proposed above model (\ref{3.3}) possesses only
cohomologically trivial solutions for $B_a$ since it yields
the equations of motion
\be \label{3.5}
{\mathrm d}B_a=0,
\ee
\be \label{3.6}
F^a+\Lambda^{ab}B_b=0.
\ee
[It is clear that (\ref{3.5}) follows from (\ref{3.6}). Moreover,
due to $F^a={\mathrm d}A^a$ and the initial supposition that constant
matrix $\Lambda^{ab}$ is non-degenerate, (\ref{3.6}) implies
that $B_a$ are exact 2-forms.] Hence $B_a$ merely represent
zero-classes in $H^2(X_D,\partial X_D,{\Bbb Z})\otimes{\Bbb R}$,
so $B_a$ cannot be a basis of this group.

Nevertheless, one can attain a cohomological non-triviality of
$B_a$ as solutions of dynamical equations, as well as a
realization of the properties (1) and (2) mentioned in the
beginning of this section. To this end it is sufficient to
add a new set of 2-forms $E^a\in C^2(X_D,{\Bbb Z})\otimes
{\Bbb R}$ to the fields used in (\ref{3.3}). Thus we propose
the following action:
$$
S_{BFE}=
\int_{X_D}\left\{\left(B_a+\Lambda_{ab}E^b\right)\wedge F^a+
\frac{1}{2}\Lambda_{ab}\left(\Lambda^{ac}B_c-E^a\right)\wedge
\left(\Lambda^{bd}B_d-E^b\right)\right\}\equiv
$$
\be \label{3.7}
\int_{X_D}\left\{\left(B_a\wedge F^a+\frac{1}{2}
\Lambda^{ab}B_a\wedge B_b\right)+\Lambda_{ab}\left(E^a\wedge F^b+
\frac{1}{2}E^a\wedge E^b\right)-B_a\wedge E^a\right\}
\ee
where $\Lambda_{ab}$ is the matrix inverse to $\Lambda^{ab}$ 
($\Lambda_{ab}\Lambda^{bc}=\delta^c_a$).

The Abelian gauge symmetries of this action (up to boundary
terms \cite{[Hor]})
\be \label{3.8}
\delta A^a={\mathrm d}\phi^a, ~ ~ \delta B_a={\mathrm d}\chi_a, 
~ ~ \delta E^a=
\Lambda^{ab}{\mathrm d}\chi_b
\ee
($\phi^a$ are 0-forms and $\chi_a$, 1-forms) combined with
equations of motion following from (\ref{3.7})
\be \label{3.9}
{\mathrm d}B_a=0={\mathrm d}E^a,
\ee
\be \label{3.10}
F^a=0,
\ee
\be \label{3.11}
E^a=\Lambda^{ab}B_b,
\ee
tell us that the space ${\cal N}$ of classical solutions (phase
space) of the $BFE$-system is
\be \label{3.12}
{\cal N}=\left[H^2(X_D,\partial X_D,{\Bbb Z})\otimes{\Bbb R}\right]
\oplus\left[H^2(X_D,{\Bbb Z})\otimes{\Bbb R}\right]
\oplus\left[H^1(X_D,{\Bbb Z})\otimes{\Bbb R}\right]
\ee
where
\be \label{3.13} \left. \begin{array}{l}
B_a\in H^2(X_D,\partial X_D,{\Bbb Z})\otimes{\Bbb R}\cong
H^2(X_D,\partial X_D,{\Bbb R}),\\ ~ \\
E^a\in H^2(X_D,{\Bbb Z})\otimes{\Bbb R}\cong H^2(X_D,{\Bbb R}),\\
~ \\
A^a\in H^1(X_D,{\Bbb Z})\otimes{\Bbb R}\cong H^1(X_D,{\Bbb R}).
\end{array} \right\} \ee
Due to the exact sequence (\ref{2.4}) the following de Rham
groups are isomorphic:
\be \label{3.14}
H^2(X_D,\partial X_D,{\Bbb R})\cong H^2(X_D,{\Bbb R}).
\ee
Taking into account this fact as well as triviality of the group
$H^1(X_D,{\Bbb R})$, we come to the phase
space (\ref{3.1}) of the $BFE$-system described by (\ref{3.7}).
Note that the equation (\ref{3.10}) means that all connections
$A^a$ are flat (1-forms $A^a$ are closed) and cohomologically
trivial since $H^1(X_D,{\Bbb R})=0$. Moreover, two sets of
closed 2-forms $B_a$ and $E^a$ defined up to exact forms may
be considered as a family of fundamental solutions of the
$BFE$-system (\ref{3.7}).\footnote{Due to the
isomorphism (\ref{3.14}) one could make no distinction between
the levels of indices of $B_a$ and $E^a$. Initially these
different levels reflected relation of these 2-forms to the
groups dual in the Poincar\'e--Lefschetz sense; below we
wouldn't like to overlook these hereditary features.} These
2-forms determine two bases of $H^2(X_D,{\Bbb R})$
related by the non-degenerate transformation (\ref{3.11})
which is nothing more than constraint equations following
from the action (\ref{3.7}).

Let us impose on the bases $B_a$ and $E^a$ the duality
condition of the type (\ref{2.8}) in the de Rham representation
({\it cf.} \cite{[DijkWitt]})
\be \label{3.15}
\frac{1}{4\pi^2}\int_{X_D}B_a\wedge E^b=\delta^b_a.
\ee
This condition fixes the gauge 1-forms $\chi_a$ introduced in
(\ref{3.8}) which automatically determine the boundary terms.
Then in this gauge the constraints (\ref{3.11}) yield
\be \label{3.16}
\Lambda^{ab}=\frac{1}{4\pi^2}\int_{X_D}E^a\wedge E^b,
\ee
and
\be \label{3.17}
\Lambda_{ab}=\frac{1}{4\pi^2}\int_{X_D}B_a\wedge B_b.
\ee
Following \cite{[FuFuUe],[FrUh]}, note that the right-hand
sides in (\ref{3.16}) and (\ref{3.17}) are rational and
integer intersection forms, respectively, and they are
represented by the mutually inverse intersection matrices
$\Lambda^{ab}$ and $\Lambda_{ab}$ in the respective bases
$E^a$ and $B_a$. In analogy with the Yang--Mills theory
\cite{[FrUh]}, the diagonal elements of the matrix $\Lambda^{ab}$
could be called topological charges. However due to the
relation (\ref{4.12}) between Euler numbers and absolute
values of the diagonal elements of the inverse matrix
$\Lambda_{ab}$, it is natural to consider as topological
charges the inverse values of the latter ones. Thus any
Abelian $BFE$-theory is characterized by the set of
topological charges $1/|\Lambda_{aa}|, ~a\in\overline{1,r}$.

These intersection forms are basic topological invariants
of pV-cobordisms of the type $X_D$ \cite{[Orlik],[EisNeum],[SavFF]}.
Thus in our $BFE$-model the coupling (cosmological) constant
matrix $\Lambda^{ab}$ is the basic topological invariant
of spacetime (of Euclidean signature) described by the 
cobordism $X_D$, and it is 
the rational intersection form in the natural basis
$E^a$, so that our $BFE$-system (\ref{3.7}) does possess the
property (1) announced in the beginning of this section.

\section{$BFE$-systems on the two-parametric family of
pV-cobordisms as a set of cosmological models}\label{s4}
\setcounter{equation}{0}

In this section we construct a set of cosmological models
with a sequence of topology changes of 3-dimensional sections
of spacetime (in Euclidean regime). Each elementary topology
change is represented by a specific pV-cobordism 
${X_D}^{(l)}_{\phantom{(}n}$
which corresponds to the decorated graph 
${\Gamma_D}^{(l)}_{\phantom{(}n}$
(the origin of two non-negative integer parameters
$n,l\in{\Bbb Z}^+$ will be explained below). On each
pV-cobordism ${X_D}^{(l)}_{\phantom{(}n}$ is defined its 
individual Abelian
$BFE$-system of the type (\ref{3.7}) which is a pure topological
`gravity' with a `cosmological term' represented by a rational
intersection matrix ${\Lambda^{(l)}_{\phantom{(}n}}^{ab}$.

\subsection{The basic set of Seifert fibered homology spheres}
\label{s4.1}

We construct the cobordisms ${X_D}^{(l)}_{\phantom{(}n}$ using 
basic structure
elements which are plumbed 4-manifolds $P(\Gamma)$ with Seifert
fibered homology spheres (Sfh-spheres) $\Sigma(a_1,a_2,a_3)$ (with
only three exceptional fibers) as boundaries:\linebreak $\partial
P(\Gamma)=\Sigma(a_1,a_2,a_3)$. Let us remind the definition of
these Sfh-spheres: $\Sigma (\underline{a}):=\Sigma(a_1,a_2,a_3)$ 
is a smooth compact 3-manifold obtained by intersecting the
complex algebraic Brieskorn surface ${z_1}^{a_1}+{z_2}^{a_2}
+{z_3}^{a_3}=0$ ($z_i \in \Bbb C _i$) with the unit 5-dimensional 
sphere $|z_1|^2+|z_2|^2+|z_3|^2=1$, where $a_1, a_2,
a_3$ are pairwise coprime integers, $a_i > 1$. There exists a
unique Seifert fibration of this manifold with unnormalized
Seifert invariants \cite{[NeumRaym]}: $(a_i,b_i)$ subject to $e(\Sigma
(\underline{a}))= \sum_{i=1}^{3} b_i/a_i =1/a$, where $a=a_1 a_2
a_3$ and $e(\Sigma (\underline{a}))$ is the well known topological
invariant of a Sfh-sphere, its Euler number.

To construct our model of universe we need a specific family of
Sfh-spheres which would be defined in following way. First,
we define the
{\it derivative of a Sfh-sphere} $\Sigma (\underline{a})$ 
as a Sfh-sphere \be \label{4.1}
\Sigma^{(1)}(\underline{a}):=\Sigma (a_1,a_2a_3,a+1)\equiv \Sigma
(a^{(1)}_1,a^{(1)}_2,a^{(1)}_3). \ee The Euler number of this
Sfh-sphere is $e(\Sigma^{(1)}(\underline{a}))=1/a^{(1)}$ where
$a^{(1)}=a^{(1)}_1a^{(1)}_2a^{(1)}_3=a(a+1)$. By induction, we
define the derivative $\Sigma^{(l)}(\underline{a})=
\Sigma(a^{(l)}_1,a^{(l)}_2,a^{(l)}_3)$ of $\Sigma(\underline{a})$
of any order $l$. In particular, there holds the recurrent
relation \be \label{4.2} a^{(l)}=a^{(l-1)}\left(a^{(l-1)}+1\right)
\ee for a product of three Seifert invariants $a^{(l)}
=a^{(l)}_1a^{(l)}_2a^{(l)}_3$. Second, we define a sequence of
Sfh-spheres which we shall call {\em primary sequence}. Let $p_i$
be the $i$th prime number in the set of the positive
integers $\Bbb N$, {\em e.g.}, $p_1=2, ~ p_2=3,\dots$. The primary
sequence of Sfh-spheres is defined as \be \label{4.3}
\{\Sigma(q_i,p_{i+1},p_{i+2})|i\in{\Bbb N}\} \ee where
$q_i=p_1\cdots p_i$. Finally, to the end of constructing our model
of universe, we include in this sequence as its first two terms
the usual {\em 3-dimensional spheres} $S^3$ {\em with Seifert
fibrations} (Sf-spheres) determined by the mappings
$h_{pq}:S^3\rightarrow S^2$ in their turn defined as
$h_{pq}(z_1,z_2)=z^p_1/z^q_2$, $p,q\in{\Bbb N}$ \cite{[Scott]}. Recall
that $S^3=\left\{(z_1,z_2) \left|
|z_1|^2+|z_2|^2=1\right.\right\}$ and $z^p_1/z^q_2 \in{\Bbb
C}\cup\{\infty\}\cong S^2$. We denote these two Sf-spheres as
$\Sigma(1,1,2)$, $p=1, q=2$ and $\Sigma(1,2,3)$, $p=2, q=3$. In
these notations we use an additional third number (unit) which
corresponds to an arbitrary regular fiber. This will enable us to
take derivatives of Seifert fibrations on $\Sigma(1,1,2)$ and
$\Sigma(1,2,3)$ by the same rule (\ref{4.1}) as for other members
of the sequence (\ref{4.3}).

\begin{table*}
\caption{ \label{tab:0} Euler numbers of Sf- and Sfh-spheres.}
\begin{center}{\footnotesize\begin{tabular}
[c]{|c|l|l|l|l|l|}\hline\hline
$_{i}\diagdown^{l}$ & 0 & 1 & 2 & 3 &
4\\\hline\hline
0&$\mathbf{0.5}$ & $0.166$ & $2.38\times10^{-2}$ & $5.53\times10^{-4}$ &
$3.06\times10^{-7}$\\\hline
1&$0.166$ & $2.38\times10^{-2}$ &
$5.53\times10^{-4}$ & $3.06\times10^{-7}$ & $9.39\times10^{-14}%
$\\ \hline 2&$3.33\times10^{-2}$ & $\mathbf{ 1.07\times10^{-3}}$ &
$1.15\times10^{-6}$ & $1.33\times10^{-12}$ &
$1.78\times10^{-24}$\\\hline 3&$4.76\times10^{-3}$ &
$2.26\times10^{-5}$ & $5.09\times10^{-10}$ & $2.59\times10^{-19}$
& $6.73\times10^{-38}$\\\hline 4&$4.33\times10^{-4}$ &
$1.87\times10^{-7}$ & $\mathbf{3.51\times 10^{-14}}$ &
$1.23\times10^{-27}$ & $ 1.52\times10^{-54}$\\\hline
5&$3.33\times10^{-5}$ & $1.11\times10^{-9}$ & $1.23\times10^{-18}$
& $1.51\times10^{-36}$ & $2.29\times10^{-72}$\\\hline
6&$1.96\times10^{-6}$ & $3.84\times10^{-12}$ &
$1.47\times10^{-23}$ & $\mathbf{2.17\times10^{-46}}$ &
$4.70\times10^{-92}$\\\hline 7&$1.03\times10^{-7}$ &
$1.06\times10^{-14}$ & $1.13\times10^{-28}$ & $1.28\times10^{-56}$
& $1.63\times10^{-112}$\\\hline 8&$4.48\times10^{-9}$ &
$2.01\times10^{-17}$ & $4.04\times10^{-34}$ & $1.64\times10^{-67}$
& $\mathbf{2.66\times10^{-134}}$\\\hline
\end{tabular} } \end{center}
\end{table*}

Now we form the family of manifolds corresponding to the first
nine primary Sfh-spheres and their derivatives up to the fourth
order, \be \label{4.4}
\{\Sigma^{(l)}(q_{i-1},p_i,p_{i+1})|i\in\overline{0,8},
l\in\overline{0,4}\}. \ee Note that the subfamily
corresponding to $i=0,1$ is built of the ordinary spheres $S^3$
with fixed Seifert fibrations. In order to include the Sf-spheres
$\Sigma(1,1,2)$ and $\Sigma(1,2,3)$ in this family, one has to put
$q_{-1}=q_0=p_0=1$. {\em E.g.}, for the well known Poincar\'e
homology sphere $\Sigma(p_1, p_2,p_3)=\Sigma(2,3,5)$, the sequence
of derivatives is
$$ \begin{array}{l}
\Sigma^{(1)}(2, 3, 5) = \Sigma(2, 15, 31),\\
\Sigma^{(2)}(2, 3, 5) =\Sigma(2, 465, 931),\\
\Sigma^{(3)}(2, 3, 5) = \Sigma(2, 432915, 865831),\\
\Sigma^{(4)}(2, 3, 5) = \Sigma(2, 374831227365, 749662454731).
\end{array}
$$

The calculation results for Euler numbers of Seifert structures of
Sf- and Sfh-spheres in the family (\ref{4.4}) are given in the
table \ref{tab:0}. We find  that for the subfamily 
\be \label{4.5}
\left\{\left. \Sigma^{(l)}_l=\Sigma^{(l)}(q_{2l-1},p_{2l},p_{2l+1})
\right|l\in\overline{0,4}\right\} \ee
the Euler numbers (the
boldface numbers) reproduce fairly well the
experimental  hierarchy of {\it dimensionless low-energy coupling}
(DLEC) constants of fundamental interactions, see table
\ref{tab:1}.

\begin{table}[h]
\caption{ \label{tab:1} Euler numbers {\it vs.} experimental
DLEC constants.}
\begin{center}
\begin{tabular}[c]{|l|l|l|l|}\hline\hline
$l$ & {$e\left(  \Sigma^{(l)}_{\phantom{|}l}\right) $} & Interaction &
$\alpha_{{\rm exper}}$\\ \hline
$0$ & $0.5$ & strong & $1$\\ \hline
$1$ & $1.07\times10^{-3}$ & electromagnetic &
$7.20\times10^{-3}$\\ \hline
$2$ & $3.51\times10^{-14}$ & weak &
$3.04\times10^{-12}$\\ \hline
$3$ & $2.17\times10^{-46}$ & gravitational &
$2.73\times10^{-46}$\\ \hline
$4$ & $2.70\times10^{-134}$ & cosmological & $<10^{-120}$\\
\hline \end{tabular}
\end{center}
~~~ \\
 {\footnotesize
{\bf Notes:} {\bf 1.} The dimensionless strong interaction
constant is $\alpha_{\mathrm{st}}= G/\hbar c$, $G$ characterizes
the strength of the coupling of the meson field to the nucleon. 
{\bf 2.} The fine structure (electromagnetic) constant
is $\alpha_{\mathrm{em}}=e^2/\hbar c$. {\bf 3.} The dimensionless
weak interaction constant is $\alpha_{\mathrm{weak}}=(G_{\mathrm{F}}/\hbar
c)(m_{\mathrm e}c/\hbar)^2$, $G_{\mathrm{F}}$ being the Fermi constant 
($m_{\mathrm e}$ is mass
of electron). {\bf 4.} The
dimensionless gravitational coupling constant is 
$\alpha_{\mathrm{gr}}=G_Nm_{\mathrm e}^2/\hbar c$, 
$G_{\mathrm{N}}$ being the 
Newtonian gravitational
constant. {\bf 5.} The cosmological constant $\Lambda$ multiplied
by the squared Planckian length is $\alpha_{\mathrm{cosm}}=\Lambda
G_{\mathrm{N}}\hbar /c^3$.  The mentioned dimensionless constants (except
the cosmological one) are also known as Dyson numbers.}
\end{table}
~~~~~\\

\noindent\underline{Observation 2.} Usually one calls as fundamental
(universal) such an interaction which is essentially characterized
by only one coupling constant. For example, the weak interaction is
universal since it is characterized only by
the Fermi constant $G_{\rm F}$ (if effects of mixing different
fundamental particles are not taken into account).
In this sense each interaction given in table 2 is really
characterized by a single coupling constant and is universal,
though only at low energies $E_{\rm low}\ll M_W\simeq 80$ GeV.
At higher energies a unification of interactions takes place, and
the collection of fundamental interactions changes. The collection
of coupling constants changes too. In our model this situation
can be found in table 3; {\it cf.} also subsection 4.2.\\

The agreement of the calculated hierarchy of the DLEC
constants and the experimental data suggests the idea to construct
a model of universe with spacelike sections obtained by splicing 
of Sf- and Sfh-spheres, see subsection 4.2 and \cite{[EfMiHer]}
(compact locally homogeneous universes with spatial sections
homeomorphic to Seifert fibrations were considered in \cite{[KoTaHo]}).
To this end we primarily have to reduce and repara\-metrize the
family (\ref{4.4}). First, in accordance with (\ref{4.5}), we
eliminate the Sf- and Sfh-spheres with odd numbers $i$ introducing
a new parameter $n\in \overline{0,4}$ related to $i$ as $i=2n$.
Then (in certain cases) it  is convenient also to use another
parameter $t=n-l, ~ t\in \overline{-4,4}$. The resulting family of
Sf- and Sfh-spheres is \be
  \label{4.6}
\begin{array}{l}
\left\{\left.\Sigma^{(l)}_n=
\Sigma\left({a_1}^{(l)}_n,{a_2}^{(l)}_n,{a_3}^{(l)}_n\right)\right|
n\in\overline{0,4},l\in\overline{0,4} \right\}:=\\\left\{\left.
\Sigma^{(n-t)}(q_{2n-1},p_{2n},
p_{2n+1})\right|
n\in\overline{0,4},t\in\overline{-4,4} \right\},
\end{array}
\ee which contains (\ref{4.5}) as a subset for $t=0$, {\em i.e.}
when $n=l$. The Euler numbers of this family of Sfh-spheres are
given in the table \ref{tab:2}.

\begin{table*}{\scriptsize
\caption{\label{tab:2}Euler number of $(n,t)$-family of
Sf- and Sfh-spheres.}
~~~~~\\
\begin{tabular}{|c|l|l|l|l|l|l|l|l|l|}
\hline\hline
$\hspace*{-.4cm}_{n}$\hspace*{-.2cm}$\diagdown$\hspace*{-.1cm}$^{t}
$\hspace*{-.4cm} & 4 & 3 & 2 & 1 &
  0 & $-1$ & $-2$ & $-3$ & $-4$\\\hline\hline
0 &  &  &  &  &\hspace*{-.2cm} $\mathbf{5.0\hspace*{-.1cm}\times 10^{-1}}$\hspace*{-.2cm}
&\hspace*{-.2cm}
$1.7\hspace*{-.1cm}\times\hspace*{-.1cm}10^{-1}$\hspace*{-.2cm}
&\hspace*{-.2cm}
$2.3\hspace*{-.1cm}\times\hspace*{-.1cm}10^{-2}$\hspace*{-.2cm}
&\hspace*{-.2cm}
$5.5\hspace*{-.05cm}\times\hspace*{-.1cm}10^{-4}$\hspace*{-.2cm}
&\hspace*{-.2cm} $3.1\hspace*{-.1cm}\times\hspace*{-.1cm}10^{-7}
$\hspace*{-.2cm}\\\hline 1 &  &  &  &\hspace*{-.2cm}
$3.3\hspace*{-.1cm}\times\hspace*{-.1cm}10^{-2}$ \hspace*{-.2cm}&
\hspace*{-.1cm}$\mathbf{1.1\hspace*{-.1cm}\times\hspace*{-.1cm}10^{-3}}
$\hspace*{-.2cm} &\hspace*{-.2cm}
$1.2\hspace*{-.1cm}\times\hspace*{-.1cm}10^{-6}$
\hspace*{-.2cm}&\hspace*{-.2cm}
$1.3\hspace*{-.1cm}\times\hspace*{-.1cm}10^{-12}$
\hspace*{-.2cm}&\hspace*{-.2cm}
$1.8\hspace*{-.1cm}\times\hspace*{-.1cm}10^{-24}$\hspace*{-.2cm}
&\hspace*{-.2cm} \\\hline 2 &  &  &\hspace*{-.2cm}
$4.3\hspace*{-.1cm}\times\hspace*{-.1cm}10^{-4}$\hspace*{-.2cm}
&\hspace*{-.2cm}
$1.9\hspace*{-.1cm}\times\hspace*{-.1cm}10^{-7}$\hspace*{-.2cm}
&\hspace*{-.2cm} $\mathbf{3.5
\hspace*{-.1cm}\times\hspace*{-.1cm}10^{-14}}$\hspace*{-.2cm} &
\hspace*{-.1cm}$1.2\hspace*{-.1cm}\times\hspace*{-.1cm}10^{-27}
$\hspace*{-.2cm} &
\hspace*{-.1cm}$1.5\hspace*{-.1cm}\times\hspace*{-.1cm}10^{-54}
$\hspace*{-.2cm} &  & \\\hline 3 &  &
\hspace*{-.1cm}$2.0\hspace*{-.1cm}\times\hspace*{-.1cm}10^{-6}
$\hspace*{-.2cm} &\hspace*{-.2cm}
$3.8\hspace*{-.1cm}\times\hspace*{-.1cm}10^{-12}$
\hspace*{-.2cm}&\hspace*{-.2cm} $1.5
\hspace*{-.1cm}\times\hspace*{-.1cm}10^{-23}$\hspace*{-.2cm}
&\hspace*{-.2cm}
$\mathbf{2.2\hspace*{-.1cm}\times\hspace*{-.1cm}10^{-46}}$\hspace*{-.2cm}
&\hspace*{-.2cm}
$4.7\hspace*{-.1cm}\times\hspace*{-.1cm}10^{-92}$\hspace*{-.2cm} &
&  & \\\hline 4 &
\hspace*{-.15cm}$4.5\hspace*{-.1cm}\times\hspace*{-.1cm}10^{-9}
$\hspace*{-.2cm} &\hspace*{-.2cm}
$2.0\hspace*{-.1cm}\times\hspace*{-.1cm}10^{-17}$\hspace*{-.2cm}&
\hspace*{-.1cm}$4.0
\hspace*{-.1cm}\times\hspace*{-.1cm}10^{-34}$\hspace*{-.2cm}
&\hspace*{-.2cm} $1.6\hspace*{-.1cm}\times\hspace*{-.1cm}10^{-67}$
\hspace*{-.2cm}&\hspace*{-.2cm}
$\mathbf{2.7\hspace*{-.1cm}\times\hspace*{-.1cm}10^{-134}}$
\hspace*{-.3cm}& &  &  & \\\hline
\end{tabular}}
\end{table*}

Parameter $t$ in our model is the discrete
cosmological `time', $t=0$ labelling the present state of the
universe where an observer can determine the DLEC constants
$\alpha^{(l)}_{\phantom{|}l}$ of the five ($l\in\overline{0,4}$)
fundamental
interactions (see table \ref{tab:1}; remember that in this case
$l=n$). The relation (\ref{4.2}) readily yields good estimates of
the DLEC constants $\alpha^{(l)}_l=
e\left(\Sigma^{(l)}_{\phantom{|}l}\right)\simeq (q_{2l+1})^{-2^l}$.
Remember that $q_{2l+1}=p_1\cdots p_{2l+1}$ is product of the
first $2l+1$ prime numbers in ${\Bbb N}$. Note that for $l=5$
there would be $\alpha^{(5)}_5\simeq 1.4\cdot 10^{-357}$ which is
too small to be identified with a certain experimentally
determined coupling constant, thus we impose the restriction
$n,l\in\overline{0,4}$ (see however \cite{[EfMiHer]} where this
restriction is lifted when the open discrete cosmological models
are described). Our hypothesis is that also in other ($t\neq 0$)
columns of table 3, the Euler numbers can be related to the
coupling constants as $e\left(\Sigma^{(l)}_{\phantom{|}n}\right)
\sim\alpha^{(l)}_{\phantom{|}n}$. In the next subsection we
discuss the realization of this hypothesis in our $BFE$-model
and introduce a new concept of $(n,l)$-preinteractions which
play the r\^ole of fundamental interactions in our purely
topological approach. In the framework of our model, any
low-energy $(n,l)$-preinteraction (which can be labeled, 
or better baptized, on the basis
of the coupling constants hierarchy, see table 2) can be traced
to its counterparts at $t\neq 0$ when one of the parameters
($n$ or $l$) not changes. {\it E.g.}, the counterparts of the
`cosmological' (4,4)-preinteraction are $(4,l)$-preinteractions
with $l\in\overline{0,4}$, and $(n,4)$-preinteractions ($n\in
\overline{0,4}$).

Though this approach leads to a hierarchy of the DLEC constants,
it yields neither a description of the spacetime structure 
of universe, nor other its features, therefore we pass to framing a more
constructive universe model in terms of 4-dimensional plumbed
cobordisms with boundary components 
glued by splicing of Sf- and Sfh-spheres.

\subsection{Construction of pV-cobordisms 
%${X_D}^{(l)}_{\phantom{(}n}$ 
and respective $BFE$-systems}\label{s4.2}

Let ${\Gamma_D}^{(l)}_{\phantom{(}n}$ be the decorated graph
shown in Figure 5.

\begin{figure}[h]     %[tbh]
%\vspace*{-2cm}
\begin{center}
\setlength{\unitlength}{1pt}
\begin{picture}(220,50)
\put(10,0){\line(1,0){40}} % between $a_{20}$ and $t_{11}$
\put(30,0){\line(0,-1){20}} % above $a_{10}$
\put(80,0){\line(1,0){40}} % between $t_{k1}$ and $t_{12}
\put(100,0){\line(0,-1){20}} % above $a_{11}$
\put(165,0){\line(1,0){40}} % between $t_{3n}$ and $a_{3n}$ 150+15
\put(185,0){\line(0,-1){20}} % above $a_{1n}$
\put(10,0){\makebox(.5,.5){$\bullet$}} % $a_{20}$
\put(30,-20){\makebox(.5,.5){$\bullet$}} % $a_{10}$
\put(30,0){\makebox(.5,.5){$\bullet$}} % $O$ FIRST
\put(100,0){\makebox(.5,.5){$\bullet$}} % $O$ SECOND
\put(185,0){\makebox(.5,.5){$\bullet$}} % $O$ THIRD
\put(50,0){\makebox(.5,.5){$\bullet$}} % $t_{11}$
\put(80,0){\makebox(.5,.5){$\bullet$}} % $t_{k1}$
\put(100,-20){\makebox(.5,.5){$\bullet$}} % $a_{11}$
\put(120,0){\makebox(.5,.5){$\bullet$}} % $t_{12}$
\put(165,0){\makebox(.5,.5){$\bullet$}} % $t_{3n}$
\put(185,-20){\makebox(.5,.5){$\bullet$}} % $a_{1n}$
\put(205,0){\makebox(.5,.5){$\bullet$}} % $a_{3n}$
\put(30,22){\makebox(.5,.5){$P\left(%
{\Gamma}^{(l)}_{\phantom{|}0}\right)$}} % FIRST
\put(100,22){\makebox(.5,.5){$P\left(%
{\Gamma}^{(l)}_{\phantom{|}1}\right)$}} % SECOND
\put(185,22){\makebox(.5,.5){$P\left(%
{\Gamma}^{(l)}_{\phantom{|}n}\right)$}} % THIRD
\put(55,-2.5){\makebox(20,5){...}} % between $t_{11}$ and $t_{k1}$
\put(125,-2.5){\makebox(20,5){...}} % to the right from $t_{12}$
\put(142,-2.5){\makebox(20,5){...}} % to the left from $t_{3n}
\put(133.5,-2.5){\makebox(20,5){..}} % additional
\put(-35,0){\makebox(20,5){$\setminus P\left(%
{\Gamma}^{{\rm ch}}_{\phantom{|}1}\right)$}} %
\put(24,-42){\makebox(20,5){$\setminus P\left(%
{\Gamma}^{{\rm ch}}_{\phantom{|}2}\right)$}} %
\put(57,-23){\makebox(20,5){$\setminus P\left(%
{\Gamma}^{{\rm ch}}_{\phantom{|}3}\right)$}} %
%\put(71,12){\makebox(20,5){$t_{k1}$}} %
\put(94,-42){\makebox(20,5){$\setminus P\left(%
{\Gamma}^{{\rm ch}}_{\phantom{|}4}\right)$}} %
\put(134,-23){\makebox(20,5){...$\setminus P(\Gamma)$...}} %
%\put(156,12){\makebox(20,5){$t_{3n}$}} %
\put(179,-42){\makebox(20,5){$\setminus P\left(%
{\Gamma}^{{\rm ch}}_{I-1}\right)$}} %
\put(230,0){\makebox(20,5){$\setminus P\left(%
{\Gamma}^{{\rm ch}}_{\phantom{|}I}\right)$}} %
\put(10,0){\circle{12}} \put(30,-20){\circle{12}}
\put(100,-20){\circle{12}} \put(185,-20){\circle{12}}
\put(205,0){\circle{12}} \put(65,0){\oval(42,14)}
\put(135,0){\oval(44,14)[l]} \put(150,0){\oval(44,14)[r]}
% \put(142.5,0){\oval(64,14)[r]}
\end{picture}\\
\vspace*{2cm}{\footnotesize Figure 5.}
\end{center}
\end{figure}

This graph corresponds to the result of plumbing of
elementary manifolds $P(\Gamma^{(l)}_m)$ with boundaries
homeomorphic to Sfh-spheres $\Sigma^{(l)}_m, ~m\in\overline{0,n}$
minus the plumbed manifolds $P(\Gamma^{\rm ch}_i)$ corresponding
to decorated chains $\Gamma^{\rm ch}_i, ~i\in\overline{1,I}$.
The notation $\setminus P(\Gamma^{\rm ch}_i)$ in the Figure 5
means subtraction of the 4-manifold $P(\Gamma^{\rm ch}_i)$ with
the boundary $L(p_i,q_i)$ from the 4-manifold 
$P(\widetilde{\Gamma}^{(l)}_n)$
where $\widetilde{\Gamma}^{(l)}_n$ is the graph 
${\Gamma_D}^{(l)}_{\phantom{(}n}$ without decoration of chains.
In other words, the graph ${\Gamma_D}^{(l)}_{\phantom{(}n}$
determine the pV-cobordism
\be \label{4.7}
{X_D}^{(l)}_{\phantom{(}n}=P({\Gamma_D}^{(l)}_{\phantom{(}n})=
P(\widetilde{\Gamma}^{(l)}_n)\setminus\bigcup_{i=1}^I
P(\Gamma^{\rm ch}_i)
\ee
between ${\Bbb Z}$-homology sphere
\be \label{4.8}
M^{(l)}_n=\Sigma^{(l)}_0\frac{~ ~ ~}{~}
\Sigma^{(l)}_1\frac{~ ~ ~}{~}\cdots\frac{~ ~ ~}{~}
\Sigma^{(l)}_n
\ee
obtained by consecutive splicing of Sfh-spheres $\Sigma^{(l)}_m,
~ m\in\overline{0,n}$, and the set of lens spaces $\bigsqcup_{i=1}^I
L(p_i,q_i)$. Thus
\be \label{4.9}
\partial{X_D}^{(l)}_{\phantom{(}n}=-\bigsqcup_{i=1}^I
L(p_i,q_i)\bigsqcup M^{(l)}_n,
\ee
see section \ref{s2.1} and \cite{[SavFF]} about the connection
between plumbing and splicing operations.

In order to calculate explicitly the intersection matrices
$\Lambda^{(l)\,ab}_{\,n}$ and $\Lambda^{(l)}_{\,n ~ ab}$,
we apply a special case of splicing (\ref{4.8}). Let us call as
$\Sigma^{(l)}_{~ m}=\Sigma({a_1}^{(l)}_{~ m},{a_2}^{(l)}_{~ m},
{a_3}^{(l)}_{~ m})$ the $m$-level Sfh-sphere with given
parameter $l$ [see (\ref{4.6}) for definitions]. We shall use 
here splicing only between Sfh-spheres
with the same (fixed) $l$ (though this is not obvious at the first
sight, it is an important point of our algorithm), hence we 
sometimes omit $l$ for 
brevity, {\it e.g.} ${a_j}^{(l)}_{~ m}=a_{jm}, ~ j=1,2,3$.
We also suppose that splicing is performed on the exceptional
fibers $S_{a_{3,m-1}}$ and $S_{a_{2m}}$, $m\in\overline{1,n}$,
{\it i.e.} we consider a special case of (\ref{4.8}) which reads
\be \label{4.10}
M^{(l)}_n=\Sigma^{(l)}_0\frac{~}{S_{a_{30}}~ ~ ~S_{a_{21}}}
\Sigma^{(l)}_1\frac{~}{S_{a_{31}}~ ~ ~S_{a_{22}}}\cdots
\frac{~}{S_{a_{3,n-1}} ~ ~ S_{a_{2n}}}
\Sigma^{(l)}_n.
\ee
This splicing diagram corresponds to the decorated graph
shown in Figure 6
which is a special case of that shown in Figure 5.

\begin{figure}[h]     %[tbh]
%\vspace*{-2cm}
\begin{center}
\setlength{\unitlength}{1pt}
\begin{picture}(220,50)
\put(10,0){\line(1,0){40}} % between $a_{20}$ and $t_{11}$
\put(30,0){\line(0,-1){20}} % above $a_{10}$
\put(80,0){\line(1,0){40}} % between $t_{k1}$ and $t_{12}
\put(100,0){\line(0,-1){20}} % above $a_{11}$
\put(165,0){\line(1,0){40}} % between $t_{3n}$ and $a_{3n}$ 150+15
\put(185,0){\line(0,-1){20}} % above $a_{1n}$
\put(10,0){\makebox(.5,.5){$\bullet$}} % $a_{20}$
\put(30,-20){\makebox(.5,.5){$\bullet$}} % $a_{10}$
\put(30,0){\makebox(.5,.5){$\bullet$}} % $0$ FIRST
\put(100,0){\makebox(.5,.5){$\bullet$}} % $0$ SECOND
\put(185,0){\makebox(.5,.5){$\bullet$}} % $0$ THIRD
\put(50,0){\makebox(.5,.5){$\bullet$}} % $t_{11}$
\put(80,0){\makebox(.5,.5){$\bullet$}} % $t_{k1}$
\put(100,-20){\makebox(.5,.5){$\bullet$}} % $a_{11}$
\put(120,0){\makebox(.5,.5){$\bullet$}} % $t_{12}$
\put(165,0){\makebox(.5,.5){$\bullet$}} % $t_{3n}$
\put(185,-20){\makebox(.5,.5){$\bullet$}} % $a_{1n}$
\put(205,0){\makebox(.5,.5){$\bullet$}} % $a_{3n}$
\put(30,12){\makebox(.5,.5){$0$}} % FIRST
\put(100,12){\makebox(.5,.5){$0$}} % SECOND
\put(185,12){\makebox(.5,.5){$0$}} % THIRD
\put(55,-2.5){\makebox(20,5){...}} % between $t_{11}$ and $t_{k1}$
\put(125,-2.5){\makebox(20,5){...}} % to the right from $t_{12}$
\put(142,-2.5){\makebox(20,5){...}} % to the left from $t_{3n}
\put(133.5,-2.5){\makebox(20,5){..}} % additional
\put(1,12){\makebox(20,5){$a_{20}$}} %
\put(24,-37){\makebox(20,5){$a_{10}$}} %
\put(44,12){\makebox(20,5){$t_{11}$}} %
\put(71,12){\makebox(20,5){$t_{k1}$}} %
\put(94,-37){\makebox(20,5){$a_{11}$}} %
\put(114,12){\makebox(20,5){$t_{12}$}} %
\put(156,12){\makebox(20,5){$t_{3n}$}} %
\put(179,-37){\makebox(20,5){$a_{1n}$}} %
\put(199,12){\makebox(20,5){$a_{3n}$}} %
\put(10,0){\circle{12}} \put(30,-20){\circle{12}}
\put(100,-20){\circle{12}} \put(185,-20){\circle{12}}
\put(205,0){\circle{12}} \put(65,0){\oval(42,14)}
\put(135,0){\oval(44,14)[l]} \put(150,0){\oval(44,14)[r]}
% \put(142.5,0){\oval(64,14)[r]}
\end{picture}\\
\vspace*{2cm}{\footnotesize Figure 6.}
\end{center}
\end{figure}

Now we calculate rational intersection forms ${\Lambda^{(l)}_n}^{ab}$
of pV-cobordisms ${X_D}^{(l)}_{\phantom{(}n}$ which correspond
to graphs ${\Gamma_D}^{(l)}_{\phantom{(}n}$ for $n,l\in\overline{0,4}$.
These matrices are obtained in the natural basis ${E^{(l)}_n}^a$ of
$H^2\left({X_D}^{(l)}_{\phantom{(}n},{\Bbb R}\right)$ which is
Poincar\'e dual to the natural basis $\left[{z^{(l)}_n}_a\right]$
of $H_2\left({X_D}^{(l)}_{\phantom{(}n},
\partial{X_D}^{(l)}_{\phantom{(}n},{\Bbb R}\right)$ defined in
\cite{[SavFF]} and briefly described in subsection \ref{s2.1}. Note
that the duality of these bases is defined 
by the pairing operation as
\be \label{4.11}
\left<{E^{(l)}_n}^a,\left[{z^{(l)}_n}_b\right]\right>=
\int_{{z^{(l)}_n}_b}{E^{(l)}_n}^a=\delta^a_b
\ee
where the relative cycles (mod $\partial{X_D}^{(l)}_{\phantom{(}n}$)
${z^{(l)}_n}_b$ are representatives of cohomological classes
$\left[{z^{(l)}_n}_b\right]$. Here rank
$H^2\left({X_D}^{(l)}_{\phantom{|}n},{\Bbb R}\right)=$ rank 
$H_2\left({X_D}^{(l)}_{\phantom{|}n},
\partial{X_D}^{(l)}_{\phantom{(}n},{\Bbb R}\right)=n+1$. Hence
the set of Abelian forms of the (\ref{3.7})-type $BFE$-system
corresponding to the cobordism ${X_D}^{(l)}_{\phantom{|}n}$, is
$\left({A^{(l)}_n}^a,{B^{(l)}_n}_a,{E^{(l)}_n}^a\right), ~a\in
\overline{0,n}$.

\noindent\underline{Observation 3.} The calculation of
${\Lambda^{(l)}_{\phantom{|}n}}^{ab}$ was given in practical
terms by Saveliev in \cite{[SavFF]}. A vertex $v$ and a
decorating oval (or circle) are called ajacent if the oval
intersects an edge one of whose endpoints is $v$. The generators
of the cohomology group
$H^2\left({X_D}^{(l)}_{\phantom{|}n},{\Bbb Z}\right)$ correspond
to the vertices of ${\Gamma_D}^{(l)}_{\phantom{|}n}$ outside of
all decorating ovals. Given such a vertex $v_a$ weighted
by an integer $e^a$, we have (see subsection 4.1 for basic
definitions)
$$
{\Lambda^{(l)}_{\phantom{|}n}}^{aa}=e^a-\sum d_i, ~ d_i=
\det A\left(\Gamma^{{\rm pch}}_i\right)/\det A\left(\Gamma^{{\rm ch}}_i\right).
$$
Here $A(\Gamma)$ is an integer intersection matrix (\ref{2.1})
corresponding to the graph $\Gamma$. The summation goes over
all indices $i$ such that $v_a$ is adjacent to the oval
containing $\Gamma^{{\rm ch}}_i$. As $\Gamma^{{\rm pch}}_i$
is denoted the portion of the chain $\Gamma^{{\rm ch}}_i$
obtained by by removing the vertex of $\Gamma^{{\rm ch}}_i$
adjacent to $v_a$ and deliting all its adjacent edges. (Note
that the determinant of an empty graph is equal to 1.) For
any two generating vertices $v_a$ and $v_b$ connected by an
edge inside ${\Gamma_D}^{(l)}_{\phantom{|}n}$ away from the
decorating ovals ${\Lambda^{(l)}_{\phantom{|}n}}^{ab}=1$.
If these two vertices are adjacent to the same decorating
oval enclosing a chain $\Gamma^{{\rm ch}}_k$, we have
$$
{\Lambda^{(l)}_{\phantom{|}n}}^{ab}=1/\det A\left(\Gamma^{{\rm 
ch}}_k\right).
$$
Note that we use unnormalized Seifert invariants 
(\cite{[NeumRaym]} and subsection 2.1), thus all generating
vertices $v_a$ have in our diagrams weights $e^a=0$, see
figure~6.

To begin with, we consider as an example the subset of rational
intersection matrices ${\Lambda^{(l)}_{\phantom{|}n}}^{ab}$ 
corresponding to
$n=l$, {\it i.e.} the matrices ${\Lambda^{(l)}_{\phantom{|}l}}^{ab}$, 
$l\in
\overline{0,4}$, $a,b\in\overline{0,l}$, calculated with the use of Saveliev's algorithm \cite{[SavFF]} (see also \cite{[EisNeum]}):
$$\left(\Lambda^{(0)}_{\phantom{(}0}\right)=
\left(\Lambda_{{\rm strong}}\right)=[\mathbf{0.5}]$$

$$\left(\Lambda^{(1)}_{\phantom{(}1}\right)=
\left(\Lambda_{{\rm elmag}}\right)=\left[
\begin{array}
[c]{cc}%
9.5\times10^{-2} & \mathbf{-1.3\times10}^{-2}\\
-1.3\times10^{-2} & 6.1\times10^{-4}%
\end{array}
\right]  $$

$$\left(\Lambda^{(2)}_{\phantom{(}2}\right)=
\left(\Lambda_{{\rm weak}}\right)=\left[
\begin{array}
[c]{ccc}%
9.8\times10^{-3} & -1.3\times10^{-4} & 0\\
-1.3\times10^{-4} & 1.6\times10^{-6} & \mathbf{-6.7\times10}^{-12}\\
0 & -6.7\times10^{-12} & 3.9\times10^{-17}%
\end{array}
\right]  $$

$$\left(\Lambda^{(3)}_{\phantom{(}3}\right)=
\left(\Lambda_{{\rm grav}}\right)=$$
$$
\left[
\begin{array}
[c]{cccc}%
1.9\times10^{-4} & -1.8\times10^{-8} & 0 & 0\\
-1.8\times10^{-8} & 1.8\times10^{-12} & -1.4\times10^{-21} & 0\\
0 & -1.4\times10^{-21} & 1.2\times10^{-27} & \mathbf{-2.2\times10}^{-40}\\
0 & 0 & -2.2\times10^{-40} & 4.1\times10^{-53}%
\end{array}
\right]  $$

$$\left(\Lambda^{(4)}_{\phantom{(}4}\right)=
\left(\Lambda_{{\rm cosm}}\right)=$$
$$\left[
\begin{array}
[c]{ccccc}%
1.0\times10^{-7} & -4.9\times10^{-16} & 0 & 0 & 0\\
-4.9\times10^{-16} & 2.4\times10^{-24} & -5.5\times10^{-41} & 0 & 0\\
0 & -5.5\times10^{-41} & 1.5\times10^{-54} & -1.2\times10^{-76} & 0\\
0 & 0 & -1.2\times10^{-76} & 4.7\times10^{-92} & 
\mathbf{-6.9\times10}^{-119}\\
0 & 0 & 0 & -6.9\times10^{-119} & 1.0\times10^{-145}%
\end{array}
\right]  $$
Note that all numbers in these matrices are rational; they are
given here up to two significant digits.

In our model, this subset of matrices is associated with the
contemporary ($t=n-l=0$) stage of universe, and with the low-energy
sector of fundamental interactions (see the boldface column
in table 3). Note that the matrix elements
${\Lambda^{(l)}_{\phantom{|}l}}^{l,l-1}={\Lambda^{(l)}_{\phantom{|}
l}}^{l-1,l}$ (boldface
numbers) are subjected to an hierarchy similar to that of the
Euler numbers $e\left(\Sigma^{(l)}_{\phantom{|}l}\right)$.
Thus we shall name the $(l,l)$-preinteraction (corresponding
to ${\Lambda^{(l)}_{\phantom{|}l}}^{ab}$) in the same way as
the fundamental interaction labelled by the same parameter $l$
in table 2. It is clear that the rational intersection matrices
${\Lambda^{(l)}_{\phantom{|}n}}^{ab}$ contain more numerical
information about $(n,l)$-preinteractions than the Euler
numbers $e\left(\Sigma^{(l)}_{\phantom{|}n}\right)$. To disentangle
this information, it is worth passing to the inverse matrices
${\Lambda^{(l)}_{\phantom{|}n}}_{ab}$ being integer intersection
matrices:
$$(\Lambda^{(0)}_{\phantom{|}4})^{-1}=$$
$$\left[ 
\begin{array}{ccccc}
\mathbf{-2} & 10 & -660 & 2.4\times 10^{6} & -1.6\times 10^{12} \\ 
10 & -30 & 1980 & -7.1\times 10^{6} & 4.9\times 10^{12} \\ 
-660 & 1980 & -2310 & 8.3\times 10^{6} & -5.7\times 10^{12} \\ 
2.4\times 10^{6} & -7.1\times 10^{6} & 8.3\times 10^{6} & -5.1\times
10^{5} & 3.5\times 10^{11} \\ 
-1.6\times 10^{12} & 4.9\times 10^{12} & -5.7\times 10^{12} & 3.5\times
10^{11} & -2.2\times 10^{8}%
\end{array}%
\right] $$

$$(\Lambda^{(1)}_{\phantom{|}4})^{-1}=$$
$$\left[ 
\begin{array}{ccccc}
-6 & 124 & -2.8\times 10^{5} & 1.4\times 10^{11} & -{3.2\times 10^{19}}
\\ 
124 & \mathbf{-930} & 2.1\times 10^{6} & -1.1\times 10^{12} & {2.4\times 10^{20}}
\\ 
-2.8\times 10^{5} & 2.1\times 10^{6} & -5.3\times 10^{6} & 2.7\times
10^{12} & -{6.1\times 10^{20}} \\ 
1.4\times 10^{11} & -1.1\times 10^{12} & 2.7\times 10^{12} & -2.6\times
10^{11} & {5.8\times 10^{19}} \\ 
-{3.2\times 10^{19}} & {2.4\times 10^{20}} & -{6.1\times 10^{20}} & {%
5.8\times 10^{19}} & -5.0\times 10^{16}%
\end{array}%
\right]  $$

$$(\Lambda^{(2)}_{\phantom{|}4})^{-1}=$$
$$\left[ 
\begin{array}{ccccc}
-42 & 11172 & -1.9\times 10^{9} & 2.2\times 10^{17} & -{2.1\times 10^{28}}
\\ 
11172 & -8.7\times 10^{5} & 1.5\times 10^{11} & -{1.7\times 10^{19}} & {%
1.6\times 10^{30}} \\ 
-1.9\times 10^{9} & 1.5\times 10^{11} & \mathbf{-2.9\times 10}^{13} & {3.2\times
10^{21}} & -{3.1\times 10^{32}} \\ 
2.2\times 10^{17} & -{1.7\times 10^{19}} & {3.2\times 10^{21}} & -{%
6.8\times 10^{22}} & {6.6\times 10^{33}} \\ 
-{2.1\times 10^{28}} & {1.6\times 10^{30}} & -{3.1\times 10^{32}} & {%
6.6\times 10^{33}} & -{2.5\times 10^{33}}%
\end{array}%
\right] $$

$$(\Lambda^{(3)}_{\phantom{|}4})^{-1}=$$
$$\left[ 
\begin{array}{ccccc}
-1806 & 7.3\times 10^{7} & -7.2\times 10^{16} & {4.0\times 10^{29}} & -{%
7.4\times 10^{45}} \\ 
7.3\times 10^{7} & -7.5\times 10^{11} & {7.4\times 10^{20}} & -{%
4.1\times 10^{33}} & {7.6\times 10^{49}} \\ 
-7.2\times 10^{16} & {7.4\times 10^{20}} & -{8.1\times 10^{26}} & {%
4.5\times 10^{39}} & -{8.3\times 10^{55}} \\ 
{4.0\times 10^{29}} & -{4.1\times 10^{33}} & {4.5\times 10^{39}} & -{%
\mathbf{4.6\times 10}^{45}} & {8.6\times 10^{61}} \\ 
-{7.4\times 10^{45}} & {7.6\times 10^{49}} & -{8.3\times 10^{55}} & {%
8.6\times 10^{61}} & -{6.14\times 10^{66}}%
\end{array}%
\right] $$

$$(\Lambda^{(4)}_{\phantom{|}4})^{-1}=$$
$$\left[ 
\begin{array}{ccccc}
-3.3\times 10^{6} & 2.7\times 10^{15} & -{8.8\times 10^{31}} & {%
1.2\times 10^{54}} & -{7.8\times 10^{80}} \\ 
2.7\times 10^{15} & -{5.6\times 10^{23}} & {1.8\times 10^{40}} & -{%
2.4\times 10^{62}} & {1.6\times 10^{89}} \\ 
-{8.8\times 10^{31}} & {1.8\times 10^{40}} & -{6.6\times 10^{53}} & {%
8.7\times 10^{75}} & -{5.9\times 10^{102}} \\ 
{1.2\times 10^{54}} & -{2.4\times 10^{62}} & {8.7\times 10^{75}} & -{%
2.2\times 10^{91}} & {1.5\times 10^{118}} \\ 
-{7.8\times 10^{80}} & {1.6\times 10^{89}} & -{5.9\times 10^{102}} & {%
1.5\times 10^{118}} & {\mathbf{-3.8\times 10}^{133}}%
\end{array}%
\right] $$
Note that all numbers in these matrices are integer; they are
given here up to two significant digits.

The properties of these matrices suggest the following observations:\\
(1) Every ${\Lambda^{(l)}_{\phantom{(}n}}_{ab}$, 
being $(n+1)\times(n+1)$-matrix,
contains the $n\times n$-matrix ${\Lambda^{(l)}_{n-1}}_{ab}$ as the
minor with respect to its last diagonal element. (This is not the
case for the rational intersection 
matrices ${\Lambda^{(l)}_{\phantom{(}n}}^{ab}$.)
Thus all homological information about the set of pV-cobordisms
$\left\{\left.{X_D}^{(l)}_{\phantom{(}n}\right|n\in\overline{0,4}
\right\}$ is contained in the $5\times5$-matrices
${\Lambda^{(l)}_{\phantom{(}4}}_{ab}$; all these matrices, 
and only they are
shown in our list.\\
(2) Absolute values of diagonal elements 
$\left|{\Lambda^{(l)}_{\phantom{(}4}
}_{nn}\right|$ of the integer intersection matrix
${\Lambda^{(l)}_{\phantom{(}4}}_{ab}$ coincide with Seifert's invariants
$a^{(l)}_{\phantom{(}n}=a^{(l)}_{1n}a^{(l)}_{2n}a^{(l)}_{3n}$
($n\in\overline{0,4}$) of Sfh-spheres $\Sigma^{(l)}_{\phantom{(}n}$.
Consequently, these diagonal elements determine the Euler
numbers of Sfh-spheres 
\be \label{4.12}
e\left(\Sigma^{(l)}_{\phantom{(}n}\right)=
1/\left|{\Lambda^{(l)}_{\phantom{(}4}}_{nn}\right|
\ee 
which in the subsection
\ref{s4.1} were related to the coupling constants 
$\alpha^{(l)}_{\phantom{(}n}\sim e\left(\Sigma^{(l)}_{\phantom{(}
n}\right)$. It is interesting that just these quantities we
juxtaposed in section \ref{s3} to the topological charges
known in the Yang--Mills theory [see the considerations
following to (\ref{3.17})]. Therefore these quantities are in
our model the most immediate analogues of coupling constants.
We remind that in section \ref{s3}, inverse ones of the matrices
${\Lambda^{(l)}_{\phantom{(}4}}_{ab}$ were called cosmological
(or coupling) constant matrices. Now this name acquires
justification not only in the framework of the generalized
$BF$-theory, but also from the viewpoint of a comparison with
the experimental low-energy coupling constants. In fact, the
boldface-written elements of the matrices
${\Lambda^{(l)}_{\phantom{(}4}}_{ab}$ correspond to coupling
constants at $t=n-l=0$ in the sense that
$\alpha^{(l)}_{\phantom{(}l}\sim e\left(\Sigma^{(l)}_{\phantom{(}
l}\right)=1/\left|{\Lambda^{(l)}_{\phantom{(}4}}_{ll}\right|$.
Hierarchically, to them are also
related the next near-diagonal elements
${\Lambda^{(l)}_{\phantom{(}4}}_{l,l-1}=
{\Lambda^{(l)}_{\phantom{(}4}}_{l-1,l}$.

These relations permit us to conclude that our rather exotic
cosmological model may be associated with the really existing universe.
To put it more strictly, the experimentally observed coupling
constants hierarchy may be determined by topological invariants
(intersection matrices) of cobordisms ${X_D}^{(l)}_{\phantom{|}n}$
which we relate to Euclidean spacetime regions describing
topology changes.

It is worth now speaking in more details about this evolutionary 
scheme (to the extent admissible in our model). Insofar as the
$BFE$-system (\ref{3.7}) does not possess local degrees of freedom,
the evolution ought to be understood as a sequence of topological
(phase) transitions resulting to changes of the set of topological
invariants $\Lambda^{(l)}_{\phantom{|}n}:=\left({
\Lambda^{(l)}_{\phantom{|}n}}^{ab},{\Lambda^{(l)}_{\phantom{|}n}}_{
ab}\right)$ as well as of $BFE$-system of forms
$S^{(l)}_{\phantom{|}n}:=\left({A^{(l)}_{\phantom{|}n}}^a,
{B^{(l)}_{\phantom{|}n}}_a,{E^{(l)}_{\phantom{|}n}}^a\right)$.
We consider the $BFE$-theory as a topological analogue of gravitation
with cosmological constant [but without other (phenomenological)
kinds of matter], thus our model is purely cosmological: One may
imagine a table similar to table 3; now each cell characterized by
the parameters $(n,t)$ or $(n,l=n-t)$, will be related to a
cosmological model which involves the respective ${\Bbb Z}$-homology
spheres $M^{(l)}_{\phantom{|}n}$ as spacelike sections,
and the $BFE$-system
$S^{(l)}_{\phantom{|}n}$. Instead of the Euler numbers $e\left(\Sigma^{(l)}_{\phantom{|}n}\right)=1/a^{(l)}_{\phantom{|}n}$,
in this new table there should appear the intersection
(cosmological constant) matrices
$\Lambda^{(l)}_{\phantom{|}n}$. The latter ones of course contain
much more numerical information than a mere set of coupling constants
$\alpha^{(l)}_{\phantom{|}n}\sim e\left(\Sigma^{(l)}_{
\phantom{|}n}\right)$. We shall call the collection
$U^{(l)}_{\phantom{|}n}=\left\{{X_D}^{(l)}_{\phantom{|}n},
S^{(l)}_{\phantom{|}n},\Lambda^{(l)}_{\phantom{|}n}\right\}$
primary $(n,l)$-universe, or $(n,l)$-preuniverse 
(reminiscent of pregeomety of John A. Wheeler).

From table 3 one can see that at $t=n-l=0$ values of the Euler
numbers $e\left(\Sigma^{(l)}_{\phantom{|}l}\right)$ (boldface
numbers) well
represent the hierarchy of the DLEC constants (see also table
2). Since the information on the Euler numbers
$e\left(\Sigma^{(l)}_{\phantom{|}n}\right)$ is contained in
the intersection matrices $\Lambda^{(l)}_{\phantom{|}n}$, it
is possible to conclude that the ensemble of $(n,l)$-preuniverses
$\left\{\left.U^{(l)}_{\phantom{|}n}\right|n=l\right\}$ should
correspond to the basic vacuum state of the present stage of
the composite universe with five $(l\in\overline{0,4})$
$BFE$-systems. It is remarkable that this ensemble of preuniverses
contains information about the hierarchy of dimensionless 
low-energy coupling constants of the real fundamental interactions
(see boldface numbers in the intersection matrices given above).
This hierarchy has in our model a purely topological origin,
and it springs up before any local degrees of freedom are
introduced. Thus the coupling constants which in most field 
theories have (semi-)phenomenological character, in our model
are topological invariants describing the global properties
of the spacetime (at least, in the Euclidean regime).

For other values of $t\in\overline{-4,4}$,
the ensembles of $(n,l)$-preuniverses 
\be \label{4.13}
\left\{\left.U^{(l)}_{\phantom{|}n}\right|n-l=t\right\}
\ee
describe the composite-universe states both of the `past'
($t<0$) and the `future' ($t>0$). Thus in our model, the
`real' universe is a superposition of $(n,l)$-preuniverses
at any fixed discrete time parameter $t$. Let us identify
the $BFE$-system $S^{(l)}_{\phantom{|}n}$ with the unique
`fundamental interaction' [$(n,l)$-preinteraction] acting
in the $(n,l)$-preuniverse. Then from the modified version
of table 3 (which we described above only verbally) it follows
that the number of $(n,l)$-preinteractions in the superposition
of $(n,l)$-preuniverses (\ref{4.13}) is growing from 1 to 5
when $t$ changes from $-4$ to 0. 
The further growth of $t$ from 0 to 4 results in decrease
of the number of $(n,l)$-preinteractions to one.
Thus in our model there exists the possibility to realize
the idea of unification of interactions, but in a rather
unusual form (instead of successive symmetry breakdowns in the
gauge theory, in our model a sequence of topology changes
takes place).

The above-described scheme corresponds to a closed model of universe.
Some details of the universe evolution with `inflationary' stages and
a possible treatment of unification ideas in 
closed and open cosmological models, are given in
\cite{[EfMiHer]} on the basis of $T_0$-discrete space approach.

\section{Conclusions}\label{s5}

Let us now summarize the basic features of our model and some pending
problems.

The ensemble of preuniverses $U^{(l)}_{\phantom{|}n}$
proposed in this paper involves $BFE$-systems
$S^{(l)}_{\phantom{|}n}$ which possess the basic characteristic
features of ordinary $BF$-systems \cite{[BlauThom]}; this means that
all physical fields may be gauged away locally. Thus the phase
spaces (sets of classical solutions) are finite dimensional
spaces [of the type of (\ref{3.1})] of zero-modes and have
a purely topological sense. But $BFE$-models also involve
the intersection matrices $\Lambda^{(l)}_{\phantom{|}n}$ as
analogues of coupling constants of fundamental interactions.
The latter ones are however described at the purely
topological level, thus we call them $(n,l)$-{\it pre}interactions.
Since the intersection matrices are basic topological invariants
of pV-cobordisms ${X_D}^{(l)}_{\phantom{|}n}$, in our version
of $BF$-theory coupling constants lose their usual phenomenological
character and acquire the status of topological invariants of
the spacetime manifold on which the $BFE$-system
$S^{(l)}_{\phantom{|}n}$ is constructed. Insofar as the
experimentally observed `running coupling constants' do depend
on local characteristics of interactions (such as energy density),
these `constants' have even in the low-energy case their values
different from those calculated in our model (see table 2).
However already at the topological vacuum level ({\it i.e.} in
the complete absence of local degrees of freedom of all fields
including the gravitational one) the information about the
hierarchy of (at least) the DLEC constants of real fundamental
interactions is contained.

This situation may be interpreted as a generalization of the
Mach principle in the sense of a determining influence of
the global topological (cosmological) characteristics of the
universe on the local properties of  universal interactions
`switched on' in this universe. It is appropriate to note
that in our model all (pre)interactions bear `cosmological
traces', that is, each preinteraction is forming a certain
preuniverse $U^{(l)}_{\phantom{|}n}$ where it is the only one
which is switched on. The spacetime topology of this
preuniverse is completely determined by the cosmological
constant matrix $\Lambda^{(l)}_{\phantom{|}n}$ representing
the rational intersection form of the pV-cobordism
${X_D}^{(l)}_{\phantom{|}n}$. The real universe involving
several interactions, is treated as a superposition of
preuniverses $U^{(l)}_{\phantom{|}n}$ with $n-l=t=$ const.
The problem still is how to determine `ordinary' fields
with their local degrees of freedom in conformity with
the topological structure of this superposition.

The preinteractions unification concept qualitatively 
differs from the usual scheme of unification accepted in
gauge theories. For example, the set of 
$(n,l)$-preinteractions found for $t=n-l$ is replaced
by another set of $(n',l')$-preinteractions. If $t\leq 0$
and $n'-l'=t-1$, the latter set contains one preinteraction
less than the former one. The number of preinteractions
decreases by a shift to the left from $t=0$ in table 3 (or 
in the analogue of this table verbally described
in subsection 4.2) as well.

Elementary pV-cobordisms ${X_D}^{(l)}_{\phantom{|}n}$ and
${X_D}^{(l)}_{\phantom{|}n+1}$ can be pasted into cobordisms
describing topology changes between ${\Bbb Z}$-homology
spheres. This is accompanied by creation and annihilation
of certain sets of disjoint lens spaces $L_{\rm out}$ and
$L_{\rm in}$. Pasting of these pV-cobordisms is naturally
performed along sets of pairwise homeomorphic lens spaces \cite{[Eff]} 
$L^{(l)}_{\phantom{|}n}\subset\partial{X_D}^{(l)}_{\phantom{|}n}$
and $L^{(l)}_{\phantom{|}n+1}\subset\partial{X_D}^{(l)}_{
\phantom{|}n+1}$ yielding the pV-cobordism
$$
{X_D}^{(l)}_{n,n+1}=-{X_D}^{(l)}_{\phantom{|}n}\bigcup
{X_D}^{(l)}_{\phantom{|}n+1}.
$$
The boundary of this cobordism,
$$
\partial{X_D}^{(l)}_{n,n+1}=-\left(M^{(l)}_{\phantom{|}n}
\bigsqcup L_{\rm in}\right)\bigsqcup
\left(M^{(l)}_{\phantom{|}n+1}\bigsqcup L_{\rm out}\right),
$$
contains both ${\Bbb Z}$-homology spheres
$M^{(l)}_{\phantom{|}n}\subset\partial{X_D}^{(l)}_{\phantom{|}n}$,
$M^{(l)}_{\phantom{|}n+1}\subset\partial{X_D}^{(l)}_{
\phantom{|}n+1}$ and sets of mutually non-homeomorphic
lens spaces $L_{\rm in}$, $L_{\rm out}$. Thus the pV-cobordism
${X_D}^{(l)}_{n,n+1}$ describes the topology change
$$
M^{(l)}_{\phantom{|}n}\bigsqcup L_{\rm in}\longrightarrow
M^{(l)}_{\phantom{|}n+1}\bigsqcup L_{\rm out}.
$$
Here one confronts, however, with the still open problem
of the junction of the $BFE$-systems
$S^{(l)}_{\phantom{|}n}$ and $S^{(l)}_{\phantom{|}n+1}$
which are defined on pV-cobordisms ${X_D}^{(l)}_{\phantom{|}n}$
and ${X_D}^{(l)}_{\phantom{|}n+1}$ respectively.

It is interesting that the intersection matrices
$\Lambda^{(l)}_{\phantom{|}n}$ always have signature $(- +\cdots +)$
for any values of $n$ and $l$. This may hint at the possibility
to construct a discrete model of a spacetime based on
Lorentz-signature lattices spanned on eigenvectors of
intersection matrices, the dimensionality of any lattice being
$n+1$. Realization of this approach should be based on a study
of the discrete phase space (\ref{3.2}) containing richer
cohomological information about the pV-cobordisms 
${X_D}^{(l)}_{\phantom{|}n}$ than the real vector space (\ref{3.1}).

Finally, note that in addition to the direct analogues of coupling
constants [that is, topological charges (\ref{4.12})] the
intersection matrices $\Lambda^{(l)}_{\phantom{|}n}$ contain a
large amount of numerical information about $(n,l)$-preuniverses,
so that these matrices could be considered as their numerical `code',
maybe (see tables 2 and 3) a `code' of our proper universe as well.
This information is encoded in the topology of the ordinary
3-sphere's `nearest relatives', namely in topology invariants
of ${\Bbb Z}$-homology spheres being spacelike sections
of spacetime manifolds.

In fact, we are greatly baffled by the strange results to which
led an application of quite an abstract and fundamental part of
mathematics, the algebraic topology, and we feel it to be appropriate
to conclude this paper with comforting and reassuring words of
Eugene P. Wigner:
``...the mathematical formulation of the physicist's often crude
experience leads in an uncanny number of cases to an amazingly
accurate description of a large class of phenomena. This shows
that the mathematical language has more to commend it than being
the only language which we can speak; it shows that it is, in a 
very real sense, the correct language''\cite{[Wig]}.

\section*{Acknowledgments}
We thank Nikolai Saveliev for kind interest to our work and for helpful advices and questions.

\end{document}